\documentclass[journal]{rmaa}

\usepackage{epstopdf}
\usepackage{textcomp}
\usepackage{multirow} 

\title{Radio Sources Embedded in the Dense Core B59, the 
``Mouthpiece'' of the Pipe Nebula}

 \author{
Sergio A. Dzib\altaffilmark{1},
Luis F. Rodr\'\i guez\altaffilmark{1,2},
Anabella T. Araudo\altaffilmark{1},
and
Laurent Loinard\altaffilmark{1}}
  \altaffiltext{1}{Centro de Radioastronom\'{\i}a y Astrof\'{\i}sica, Universidad
Nacional Aut\'onoma de M\'exico, Campus Morelia} 

\altaffiltext{2}{Astronomy Department, Faculty of Science, King Abdulaziz University, P.O.
Box 80203, Jeddah 21589, Saudi Arabia}

\fulladdresses{
\item Anabella T. Araudo, Sergio A. Dzib, Laurent Loinard, Luis F. Rodr\'\i guez: Centro de Radiostronom\'{\i}a
y Astrof\'\i sica, 
UNAM, A. P. 3-72, 
(Xangari), 58089 Morelia, Michoac\'an, M\'exico
(a.araudo, s.dzib, l.loinard, l.rodriguez@crya.unam.mx).
}

\shortauthor{Dzib et al.}
\shorttitle{Radio Sources in B59}

\SetVolume{50} \SetFirstPage{001} \SetYear{2013}

\resumen{Presentamos observaciones de continuo realizadas con 
el Very Large Array a 8.3~GHz del n\'ucleo denso B59 en la nebulosa de la 
Pipa. Detectamos seis fuentes compactas, de las cuales cinco est\'an asociadas 
con las cinco fuentes m\'as luminosas a 70~$\mu$m en la regi\'on, mientras que 
la fuente restante es probablemente extragal\'actica. Proponemos que 
la emisi\'on en radio es libre-libre de los vientos ionizados presentes en 
estas protoestrellas. Discutimos el impacto cinem\'atico de estos vientos en 
la nube. Tambi\'en proponemos que estos vientos son \'opticamente gruesos en 
radio pero \'opticamente delgados en rayos X y esta caracter{\'\i}stica puede 
explicar 
porque los rayos X de la magnetosfera son detectados en tres de las fuentes, 
mientras que la emisi\'on de radio est\'a probablemente dominada por emisi\'on 
libre-libre de las capas externas del viento. 
}
 
\abstract{We present Very Large Array continuum observations made at 8.3 GHz toward the dense core
B59, in the Pipe Nebula. We detect six compact sources, of which five are associated
with the five most luminous sources at 70 $\mu$m in the region, while the remaining one
is probably a background source. We propose that the radio emission is free-free from
the ionized outflows present in these protostars. We discuss the kinematical impact 
of these winds in the cloud. We also propose that these winds are optically thick
in the radio but optically thin in the X-rays and that this characteristic can
explain why X-rays from the magnetosphere are detected in three of them, while
the radio emission is most probably dominated by the free-free emission
from the external layers of the wind.
}
      

\begin{document}

\maketitle

\section{Introduction}

The Pipe Nebula is a nearby ($\sim$130 pc), massive 
($\sim10^4~M_\odot$) molecular cloud 
(Lombardi et al. 2006; Onishi et al. 1999) in
the southern edge of the Ophiuchus constellation that extends over 
$\sim6^\circ$ in the plane of the sky. Onishi et al. (1999)
identified 14 C$^{18}$O cores within the Pipe Nebula.
Of these, only the core associated with B59 (the ``mouthpiece''
of the Pipe Nebula) is known to have
active star formation. Onishi et al. (1999) reported 
a CO outflow toward the center of the B59 core. There are several
H$\alpha$ emission line stars in or near the B59 core
(Cohen \& Kuhi 1979; Herbig \& Bell 1988; Kohoutek \& Wehmeyer 2003; Herbig 2005).
Conclusive evidence of star formation
was provided by Brooke et al. (2007) who, as part of the 
Spitzer c2d Survey of Nearby Dense Cores,
detected an embedded cluster with at least 20 candidate low-mass young stars. 
More than 130 dust extinction cores have been identified
in the Pipe Nebula and only B59 shows obvious signposts of star formation
(Forbrich et al. 2009).
There are now observations of B59 in several wavelengths and in this
paper we present the analysis of sensitive, high angular resolution
radio continuum observations
made at 3.6 cm to complement our understanding of this region.

\section{Observations}

The deep infrared extinction map of Rom\'an-Z\'u\~niga et al. (2009)
shows that the densest part of the B59 core has irregular shape and
extends over $\sim5'$. Thus, a single pointing with the Very Large Array 
(VLA) of the National Radio Astronomy Observatory\footnote{The NRAO
is operated by Associated Universities
Inc. under cooperative agreement with the National Science Foundation.}
at 3.6 cm is sufficient to cover the densest part of B59.
In the archives of the VLA we found unpublished observations made in 2004 February 12 under the
project AA290 at 3.6 cm in the C configuration. These data 
were edited and calibrated using the software package Astronomical Image
Processing System (AIPS) of NRAO. The flux calibrator was 1331+305, with an
adopted flux density of 5.23 Jy. The gain calibrator was 1626$-$298,
with a bootstrapped flux density of 1.749$\pm$0.003 Jy.

We detected a total of six sources in the field. These sources are shown in
Figure 1 and their parameters are listed in Table 1.

\begin{table*}[htbp]
\footnotesize
  \setlength{\tabnotewidth}{2.0\columnwidth} 
    \tablecols{5} 
\small
  \caption{Radio Continuum 3.6 cm Sources in the B59 Dense Core}
    \begin{center}
	\begin{tabular}{lcccc}\hline\hline
	&\multicolumn{2}{c}{Position$^a$} & Flux &  \\
	\cline{2-3} 
	No. &  $\alpha$(J2000) & $\delta$(J2000) & Density$^b$ (mJy) & Counterparts$^c$ \\ 
	\hline
	1 & 17 11 13.43 &  $-$27 27 40.1 & 0.33$\pm$0.06 & Extragalactic?  \\
	2 & 17 11 17.26 &  $-$27 25 08.6 & 0.28$\pm$0.05 & 2MASS J17111726-2725081, [BHB2007] 7  \\
	3 & 17 11 21.55 &  $-$27 27 42.0 & 0.23$\pm$0.04 & 2MASS J17112153-2727417, [BHB2007] 9, 2XMMi J171121.5-272741 \\
	4 & 17 11 22.16 &  $-$27 26 01.9 & 0.24$\pm$0.04 & [DCE2008] 023, [BHB2007] 10 \\
	5 & 17 11 23.11 &  $-$27 24 32.4 & 0.63$\pm$0.05 & 2MASS J17112318-2724315, [BHB2007] 11, 2XMMi J171123.0-272432 \\
	6 & 17 11 27.00 &  $-$27 23 47.9 & 0.26$\pm$0.05 & 2MASS J17112701-2723485, [BHB2007] 13, 2XMMi J171127.0-272348 \\
	\hline\hline
\tabnotetext{a}{Positional error is estimated to be $0\rlap.{''}5$.}
\tabnotetext{b}{Total flux density corrected for the primary beam response.}
\tabnotetext{c}{2MASS = Cutri et al. (2003); [BHB2007] = Brooke et al. (2007); 2XMMi = 
Watson et al. (2009), see also http://xmmssc-www.star.le.ac.uk/Catalogue/xcat\_public\_2XMMi-DR3.html;  
[DCE2008] = Dunham et al. (2008).} 
  \label{tab:1}
\end{tabular}
\end{center}
\end{table*}

\begin{figure*}
\centering
\includegraphics[scale=0.5, angle=0]{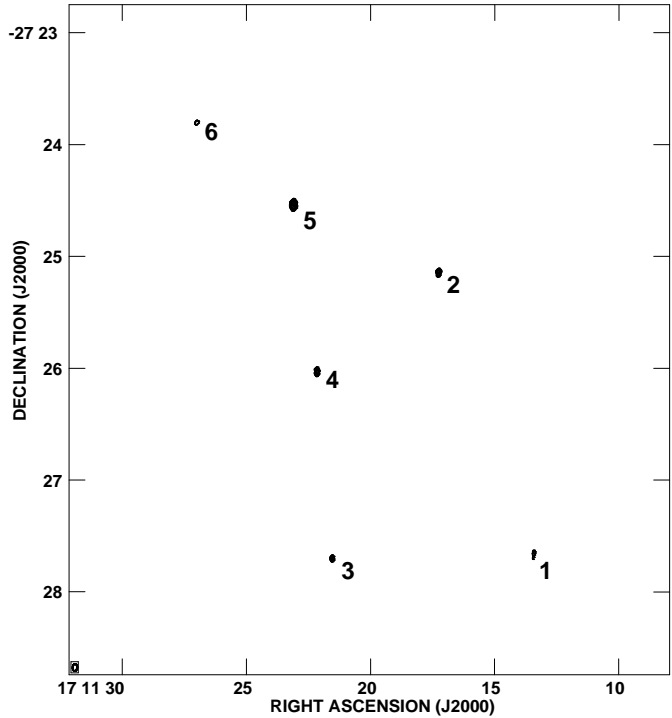}
 \caption{VLA contour image of the 3.6-cm continuum emission toward
B59. The image was made with natural weighting to optimize
the sensitivity of the image. Contours are -5, 5, 6, 8, 10, 12, 15, 20, and 30 
times 14.5 $\mu$Jy, the rms noise of the image. 
The synthesized beam, shown in the bottom left corner,
has half power full width dimensions of
$3\rlap.{''}99 \times 2\rlap.{''}34$, 
with the major axis at a position angle of $-3^\circ$. 
The six sources detected are numbered as in Table 1.
}
  \label{fig1}
\end{figure*}

\section{Discussion}

\subsection{Association of the Radio Sources with Embedded Stars}

The six sources detected have flux densities of 0.23 mJy or more.
In a solid angle of about 30 arcmin$^2$ we expect \sl a priori \rm
0.5 background sources (Windhorst et al. 1993; Fomalont et al. 2002) above 0.23 mJy. Then, 
we can safely conclude
that most of the sources are associated with B59, with perhaps one
of them being a background source. This statistical estimate is corroborated
by the existence of infrared counterparts to five of the radio sources:
2, 3, 4, 5, and 6. Remarkably, these five radio sources are spatially associated
with the five most luminous sources at 70 $\mu$m detected by Brooke et al. (2007)
and located within 0.1 pc of the peak of the molecular emission.
Three of the radio sources are associated with XMM X-ray sources (see Table 1).
As a consequence, we propose that VLA 1 is a background extragalactic
source, while the remaining five VLA sources (numbers 2 to 6)
are associated with the five most luminous sources at 70 $\mu$m
(corresponding, respectively, to the sources 7, 9, 10,
11, and 13 of Brooke et al. 2007). The brightest of the radio sources (VLA 5)
is also the youngest, being classified as class 0/I (Brooke et al. 2007), and it
is also detected at 1.3 mm (Reipurth et al. 1996).
Finally, it appears to be the exciting source of the best defined CO outflow in the
region (Duarte-Cabral et al. 2012).

\subsection{The Nature of the Radio Emission}

Low mass young stars can be radio continuum emitters by two main
mechanisms. Thermal free-free emission can originate in the partially
ionized circumstellar material, usually tracing
the base of outflowing gas. On the other hand, non-thermal
gyrosynchrotron radiation can be produced in the magnetically-active
corona of some young stars.

These two types of radiation can be differentiated observationally by their
particular characteristics. Thermal free-free radiation is
usually steady or with slow time variations (with timescales of years),
with a spectral index $\,\gtrsim\,-$0.1, and with no polarization.
In contrast, gyrosynchrotron radiation can show variation timescales
of hours to days and spectral indices ranging between -2 and 2.
It can also show significant levels of circular polarization 
(e.g. G\'omez et al. 2008). 

Unfortunately, the data available is at only one epoch and at one
frequency, so most of the criteria to distinguish between both
mechanism cannot be applied. We searched for circular polarization in
the sources, setting 3-$\sigma$ upper limits of $\sim$0.03 mJy at the
center of the image. This implies typical upper limits of 5--10\% for
the circular polarization. This lack of circular polarization favors,
although not in a definitive way, a free-free nature for the emission.
We will tentatively assume that the emission is indeed free-free mostly
on the basis of two arguments. The first
is the association of the radio sources with the
brightest 70 $\mu$m sources. While the strengh of the gyrosynchrotron
is not known to be correlated with the stellar luminosity, it is
well known that the free-free radio continuum is indeed correlated
with the stellar luminosity (i.e. AMI Consortium: Scaife
et al. 2011). The radio luminosity of
the sources detected in B59 are indicative of stars with a bolometric
luminosity of $\sim$0.5 $L_\odot$ (AMI Consortium: Ainsworth et al. 2012),
and this is consistent with the values found for the embedded stars
(Riaz et al. 2009; Covey et al. 2010).
The second argument is that the stellar sources are mostly classes 0 to II
(Brooke et al. 2007),
where gyrosynchrotron emission is rare, being present 
more frequently in the more evolved class III sources (Dzib et al. 2013, in
preparation).

\subsection{Effect of the Stellar Outflows on the Cloud}

Under this assumption, we can use the correlation between centimeter
radio continuum and molecular momentum rate
(e.g. Rodr\'\i guez et al. 2008) to estimate if the outflows been traced
by the radio continuum can play a role in the disruption of
the B59 molecular core. We obtain that each star injects on the
average a momentum rate of $8 \times 10^{-6}~M_\odot~yr^{-1}~km~s^{-1}$, for a
total of $4 \times 10^{-5}~M_\odot~yr^{-1}~km~s^{-1}$ adding the
effect of the five stars.

The mass of the nuclear region of B59 is estimated from
a near-IR dust extinction map to be 18.9 $M_\odot$ 
(Rom{\'a}n-Z{\'u}{\~n}iga et al. 2012). To accelerate this
mass to velocities of order 1 km s$^{-1}$,
we need a timescale of $\sim 5 \times 10^{5}$ years
for the winds implied by the radio
observations. This acceleration is expected to disrupt or at least significantly perturb the cloud.
Covey et al. (2010) estimate a median stellar age of 
2.6$\pm$0.8 Myr for B59. We then expect the cloud to 
be perturbed by the outflows. Duarte-Cabral et al. (2012) have
studied the molecular kinematics of the cloud and 
conclude that most or it consists 
of cold and quiescent material, mostly gravitationally bound, 
with narrow line widths. However, they also conclude
that the impact of the outflows is observed close to the
protostars as a localized increase of both C$^{18}$O line widths 
from $\sim$0.3 km s$^{-1}$ to $\sim$1 km s$^{-1}$, and 
$^{13}$CO excitation temperatures by 2-3 K. It is possible that a
significant fraction of the stellar wind's momenta has gone into
slowing the collapse of the region more than in feeding expansion
motions. It is also possible that the wind's momenta goes into
accelerating a small fraction of the gas to high velocities, while
leaving most of the cloud relatively unperturbed.
Duarte-Cabral et al. (2012) estimate that the total momentum flux of the outflows
in the cloud is $2.7 \times 10^{-4} ~M_\odot~yr^{-1}~km~s^{-1}$, about five times larger
than the total momentum flux estimated from the radio continuum. This difference
is probably due to the uncertainties involved in the estimates, but it could also
suggest that the stellar winds were stronger in the past, since the outflows are
the fossil record of the effect of the winds on the molecular cloud.

\subsection{X-ray Emission}

Of the five radio sources associated with young stars, we find that
three have also associated X-ray emission detected with XMM (see Table 1).
According to the Guedel-Benz relation (Guedel 
\& Benz 1993) there is
an intimate connection between the nonthermal, energetic electrons causing 
the gyrosynchrotron radio emission and the bulk plasma of the 
corona responsible for thermal X-rays. In other words, the presence
of X-ray emission is expected to correlate with radio non-thermal
emission. However, previously we argued  
that the observed radio emission has a thermal nature.

A possible explanation for this situation is that the X-ray emission from the 
corona can penetrate the stellar wind
and is detected by an
external observer, while the non-thermal radio emission
produced in this same region is absorbed by the free-free opacity
of the stellar wind. Then, in the radio we detect only the
thermal (free-free) emission from the wind itself. 

To test this possible explanation, we estimate the radio and
X-ray opacities expected for the winds of these young stars.
The opacity along a line of sight that points from the observer
to the center of the star is given by

$$\tau_\nu(R) = \int^\infty_R \kappa_\nu dr,$$

\noindent where $\kappa_\nu$ is the absorption coefficient,
$dr$ is the increment of pathlength, and $R$ is the radius,
with respect to the center of the star, of the point considered.

At radio wavelengths, the absorption coefficient of the free-free
process is given by (Panagia \& Felli 1975)

$$\Bigg[{{\kappa_\nu} \over {cm^{-1}}}\Bigg] = 1.06 \times 10^{-25}~\Bigg[{{T_e} \over {10^4~K}}\Bigg]^{-1.35}~
\Bigg[{{\nu} \over {GHz}}\Bigg]^{-2.1}$$ 
$${\times}~\Bigg[{{n_e} \over {cm^{-3}}}\Bigg]^2.$$

The electron density is related to the parameters of the stellar wind by

$$n_e(r) = {{\dot M} \over {4 \pi~ r^2~ V_\infty~ \mu~ m_H}},$$

\noindent where $\dot M$ is the mass loss rate, $r$ is the radius of
the point considered, $V_\infty$ is the terminal velocity
of the wind, $\mu$ is the mean atomic weight per electron, and
$m_H$ is the mass of the hydrogen atom.
Assuming that hydrogen and helium are once ionized in the wind and following
the treatment of Panagia \& Felli (1975), we find that the radius at
which $\tau_\nu(R_{thick}) = 1$, that approximately defines the
region inside which the wind is optically thick, is
given by

$$\Bigg[{{R_{thick}(\nu)} \over {R_\odot}} \Bigg] = 8.7 \times 10^2~\Bigg[{{T_e} \over
{10^4~K}}\Bigg]^{-0.45}~\Bigg[{{\nu} \over {GHz}}\Bigg]^{-0.7}$$
$$\times \Bigg[{{\dot M} \over {10^{-8}~M_\odot~yr^{-1}}}\Bigg]^{2/3}~
\Bigg[{{\mu} \over {1.2}}\Bigg]^{-2/3}~
\Bigg[{{V_\infty} \over {100~km~s^{-1}}}\Bigg]^{-2/3}.$$

Assuming an isotropic wind with a terminal velocity of 200 km s$^{-1}$
(the escape velocity of a star with a mass of 0.5 $M_\odot$ and
a radius of 4 $R_\odot$),
$\mu$ = 1.2, $T_e$ = 10$^4$ K,
and an 8.3 GHz flux density of 0.3 mJy, that is located at a
distance of 130 pc, using the formulation of Panagia \& Felli (1975),
we derive a mass loss rate of $\dot M = 1.2 \times 10^{-8}~M_\odot~yr^{-1}$.
We then derive

$$R_{thick}(8.3~GHz) \simeq 140~R_\odot.$$

This result implies that any emission from the stellar magnetosphere will
not be detected at radio wavelengths.


In the X-rays, the most important process to attenuate the emission is photoelectric
absorption. To account for this effect on $\kappa_{\nu}$ it is necessary to
know the abundances of different (metallic) species. In the present paper
we  fit as 
$\kappa(E) = 50 (E/keV)^{-2}$ cm$^{-2}$ g$^{-1}$ the dependence published by
Leutenegger et al. (2010; their Fig.~7)\footnote{In this case, $\kappa(E)$
is assumed to be constant throughout the wind. However, this assumption
is relaxed in recent studies of the $\zeta$-Puppis wind (Herv\'e et al. 2013).}.
This fit is valid for $E \geq$ 0.6 keV.
Integrating the optical depth from the observer 
up to a distance $R_{\rm thick}$ where $\tau = 1$, we obtain that 
%
$$
\left[\frac{R_{\rm thick}(X)}{R_{\odot}}\right] = 3.6 
\left[\frac{\dot M}{10^{-8}~ {\rm M_{\odot}\,yr^{-1}}}\right]
\left[\frac{V_{\infty}}{100 \,{\rm km\,s^{-1}}}\right]^{-1}
$$
$$
\times
\left[\frac{E} {\rm keV}\right]^{-2}.
$$
%
Using that $\dot M = 1.2 \times 10^{-8}~M_\odot~yr^{-1}$ and 
$V_{\infty} = 200$~km~s$^{-1}$ we derive that
at $E$ = 1 keV, 

$$R_{\rm thick}(X) \simeq  2~R_{\odot}.$$

During its main accretion phase, a solar-mass protostar is expected to have
a radius of about $4\,R_{\odot}$ (Masunaga \& Inutsuka 2000). Thus,
we find that
at $E \gtrsim 1$~keV the wind 
is optically thin, allowing us to detect the X-rays produced in the
magnetosphere.

This interpretation can be tested by observations in the sub-mm range, that in
principle will penetrate the wind and allow the detection of non-thermal radiation
at those short wavelengths.



\section{Conclusions}

In this paper we studied the radio emission from compact sources in the B59 dark core.
We report the detection of six sources at 8.3 GHz. We propose that one of these sources
is probably a background object, while the other five are associated with the five
most luminous sources at 70 $\mu$m in the region (Brooke et al. 2007).
We argue that the radio emission from these five sources is most probably
free-free from their ionized winds. 
We propose that these outflows have significantly affected the kinematics of
the core, although this effect seems to be concentrated close to the protostars.
Finally, we discuss why three of the five sources have detected X-ray emission
but apparently lack non-thermal radio emission. 
We propose that the winds are optically thin at X-rays, but optically thick in the radio,
and that this explain the detection of X-rays from the magnetosphere, while the
radio emission is dominated by the optically-thick thermal emission from the wind.
Interestingly, even when the emission
mechanisms of the X-rays and radio emission in these stars are
independent, they still fit the Guedel \& Benz (1993) relation, suggesting that
stellar winds may be a significant source of contamination in this
relation.


\acknowledgments
We are thankful to an anonymous referee for valuable comments that
improved the paper.
LFR and LL are thankful for the support
of DGAPA, UNAM, and of CONACyT (M\'exico).
This publication makes use of data products from the Two Micron All Sky Survey, which is 
a joint project of the University of Massachusetts and the Infrared Processing and 
Analysis Center/California Institute of Technology, funded by the National 
Aeronautics and Space Administration and the National Science Foundation.
This research has made use of the SIMBAD database, 
operated at CDS, Strasbourg, France.

\vskip0.5cm


\end{document}